\documentclass{article}

\usepackage[authoryear, round]{natbib}

\usepackage[T1]{fontenc}
\usepackage[utf8]{inputenc}
\usepackage{geometry}
\usepackage{float}
\usepackage{subfigure}
\usepackage{amsmath}
\usepackage{multirow}
\usepackage{booktabs}
\usepackage{mathtools}
\usepackage{authblk}
\usepackage{graphicx}
\usepackage{lscape}
\usepackage{indentfirst}
\usepackage{enumerate}
\usepackage{enumitem}
\usepackage{textcomp}
\usepackage{tabu}
\usepackage{geometry}
\usepackage{pdflscape}
\usepackage{supertabular}
\usepackage[stable]{footmisc}
\usepackage{subfigure}
\usepackage{scrextend}
\usepackage{lipsum} 
\usepackage[colorlinks=true, allcolors=black]{hyperref}
\deffootnote[0em]{0em}{0em}{\thefootnotemark\space}
\usepackage{lineno}
\usepackage{setspace}
\usepackage{appendix}
\newcommand{\beginsupplement}{%
        \setcounter{page}{1}
        \setcounter{section}{0}
         \setcounter{figure}{0}
        \renewcommand{\thesection}{S\arabic{section}}
        \renewcommand{\thefigure}{S\arabic{figure}}%
     }

\title{Population size and centrality effects on NO$_2$ air pollution across and within European cities}

\author[1]{Yufei Wei}
\author[2]{Rémi Lemoy}
\author[1]{Geoffrey Caruso\thanks{Corresponding author: geoffrey.caruso@uni.lu}}
\affil[1]{Department of Geography and Spatial Planning, University of Luxembourg, Esch-sur-Alzette, Luxembourg}
\affil[2]{UMR 6266 CNRS IDEES, University of Rouen, Mont-Saint-Aignan, France}
\date{}

\begin{document}
\maketitle
\doublespacing

\begin{abstract}
The concentration of nitrogen dioxide (NO$_2$) in ambient air is a major threat to the health of urban populations. However, the variation of NO$_2$ levels within cities and its relation to population size are still unclear. We quantify the effect of centrality and population size on NO$_2$ in 378 European cities using monitoring stations and satellite data, both giving consistent results. NO$_2$ concentration increases with population $N$ as a power law with an exponent between 0.13 and 0.22, and decreases with the distance $r$ from the center as a power law with an exponent between -0.11 and -0.18. The effects of the local context for monitoring stations data and of the regional background level for satellite measurements are important but the centrality and population size effects, when jointly estimated, are robust to these additions. Our results are summarized in a simple equation where the total air pollution over a city is proportional to $(N/r)^{0.16}$. We discuss this result and show that the benefits or dis-benefits of larger cities (scaling behavior) depend on the extent to which pollution is integrated and how the urban fringe is defined.
\end{abstract}

\newpage
\section{Introduction}
Nitrogen dioxide (NO$_2$) is a major air pollutant \citep{whoSelectPol}, and road traffic is one of its primary sources \citep{eea2016no2}. NO$_2$ is detrimental to human health, causing diseases such as lifelong asthma \citep{10.2307/20486137}, heart attacks \citep{Young2019} and cancer \citep{10.1289/ehp.1408882}. It is also a major precursor to acid rain \citep{IRWIN198829}, and is involved in the formation of photochemical smog \citep{WANG2021113319}. NO$_2$ levels above the World Health Organisation guidelines have been associated to premature deaths, e.g. over 70,000 per year across 41 European countries \citep{EEA2023}.

As human activities and transport are concentrated in cities, so is air pollution. NO$_2$ constitutes the second most harmful source of urban air pollution after particulate matter \citep{EEA2023,sicard2021urban}. In comparison to particulate matter, NO$_2$ can travel over longer distances, which means that its spatial distribution is relatively less influenced by the local context and road network geometry, and more by the overall layout and size of the city. In addition to the local aspects that differentiate places within cities, the total amount of urban activity, hence the population of cities, relative to its spatial extent (i.e. how far the city spreads out) is thus also important in the case of NO$_2$.

One of the most important dimensions for assessing the overall shape of a city is its radial profile, that is how population density or the intensity of land development decline as you move away from central areas. The radial profiles of density and artificial land are known to be similar (homothetic) across city sizes \citep{LemoyCaruso2020, Lemoy2021} and to result from residential choice \citep{alonso1964location, delloye2020alonso, Cottineauetal}. When traffic-related air pollution is added to this choice theory, \citet{SCHINDLER201712} found that pollution levels decrease with distance from the city center in line with density profiles. They also show that when households are willing to reduce their own pollution exposure, they move to the periphery. Hence density profiles become flatter when health awareness about pollution is higher, while total emissions increase, which goes against sustainability goals. In addition to traffic, air pollution also arises from energy use for heating. Although this effect has not been internalized in urban theory, density is expected to further increase pollution concentrations, again leading to a movement away from city centers and from larger cities. These are important mechanisms that must be well understood and quantified since they eventually oppose global environmental goals, which favor density and large cities, to individual health objectives, which favor peripheral locations and smaller cities. 

Empirical evidence on the effect of centrality on NO$_2$ has been found for few case studies \citep{HIEN2020134635, Kirby98, NICHOLASHEWITT1991429, WANG2020117470, BORCK2021103596} or included within multi-factorial analyses across many cities \citep{iungman2024impact}. Some research has also been done on how aggregate pollution levels scale with city size \citep{BRANIS20121110, 10.3389/fenvs.2022.844479, ijerph16091487, Lamsal2013, doi:10.1021/es103866b}. Unfortunately, none have estimated distance and city size effects simultaneously and for a large set of cities so that a more general understanding of NO$_2$ distribution within and across cities can be reached. Our goal is to fill this gap by studying a large set of European cities and to determine the influence of both city size and centrality within each city on the distribution of annual average NO$_2$ concentrations.

NO$_2$ concentration can be measured precisely at the ground level from monitoring stations or more ubiquitously within a tropospheric column using satellite sensors \citep{poPaperAirQ}. Given both measurement types have been shown to correlate only weakly \citep{Blond2007,ijerph14111415}, and even less in urban environments \citep{WALLACE20093596}, we use both and compare results.

Our analysis also contributes an example of a further integrated urban science \citep{10.2307/j.ctt9qf7m6}  through the use of functionally defined urban areas \citep{fuaIntro,Dong2024} and by explicitly relating city size (or scaling) effects to intra-urban patterns. Our approach combines inter-urban analysis, through population size, and intra-urban analysis through a radial lens, i.e. the distance to the center. We show that controlling for this internal distance rather than using mean aggregates per city is key to determining scaling behaviors. 

\section{Data and Methods}\label{Methods}
We analyze how intra-urban levels of NO$_2$ can be explained by city size and internal centrality. City size is measured as the total population of a functional urban area and denoted by $N$. Centrality is measured as the Euclidean distance from the city center, defined as the city hall, and is denoted by $r$. In this section, we present our data and the general models we use to estimate the effect of $N$ and $r$.

\subsection{Urban areas}\label{urban data}
We consider the 378 largest Functional Urban Areas (FUAs) across 33 European countries.

The FUAs, as defined by the OECD \citep[Organization for Economic Cooperation and Development,][]{fuaIntro}), consist of cities, made of dense contiguous urbanized areas \citep{cityDef} plus their surrounding commuting areas, where the percentage of people employed in the city is at least 15\% \citep{commuZoneDef}. FUAs are relevant homogeneous urban domains for comparative analyses and for the formulation of transportation and planning policies \citep{OECDCiWorUrba}. The use of comparable and functional definitions that represent daily interactions has long been recommended to better integrate knowledge in urban science \citep[again recently by][]{Dong2024} and is especially important in the case of NO$_2$ since it arises from transport, which cannot be well appreciated if only core areas or street scales are considered.
 
\begin{figure}[htbp]
\centering
\includegraphics[trim = 45mm 0 45mm 0, clip, width=0.49\textwidth]{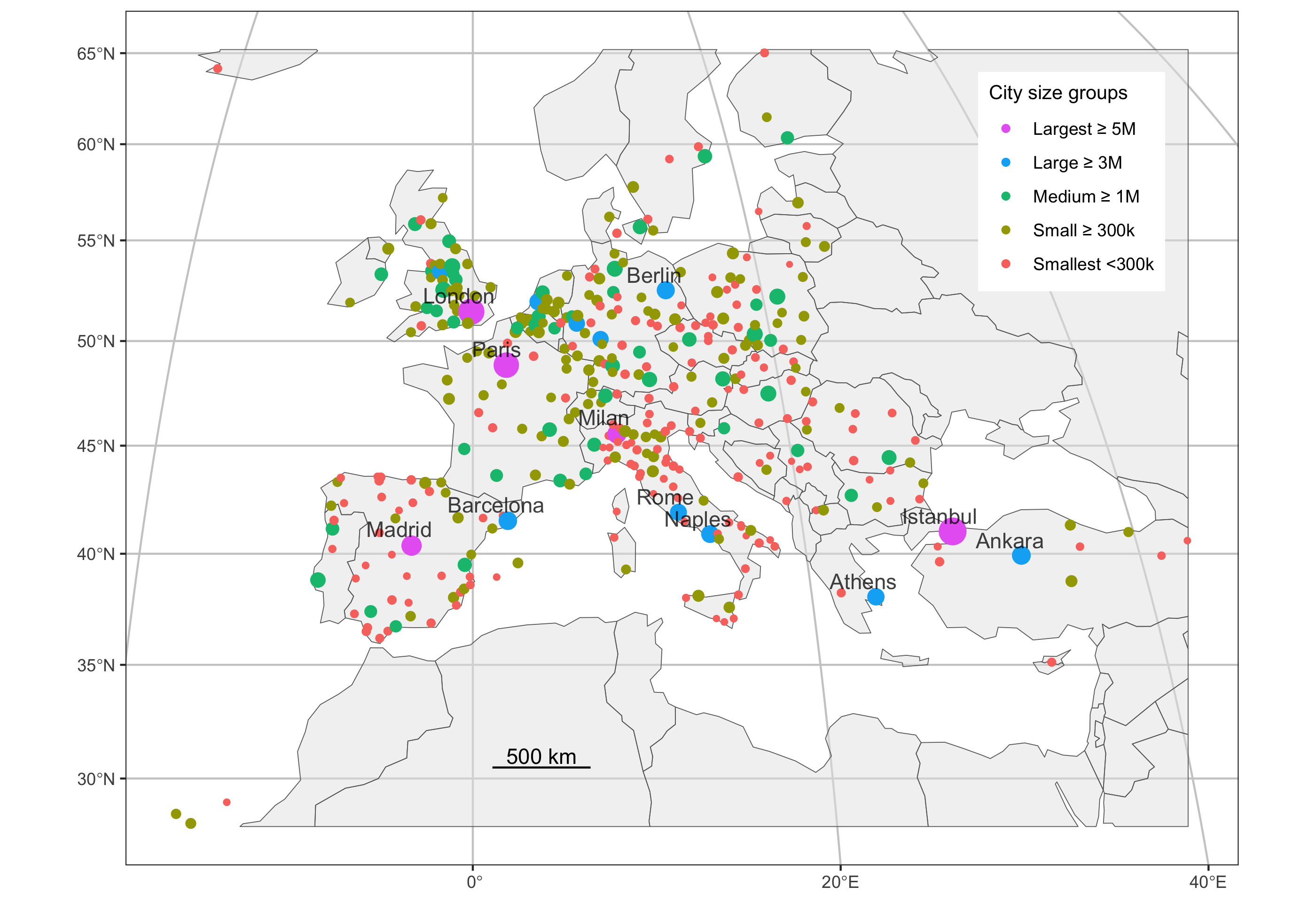}%
\includegraphics[trim = 45mm 0 45mm 0, clip, width=0.49\textwidth]{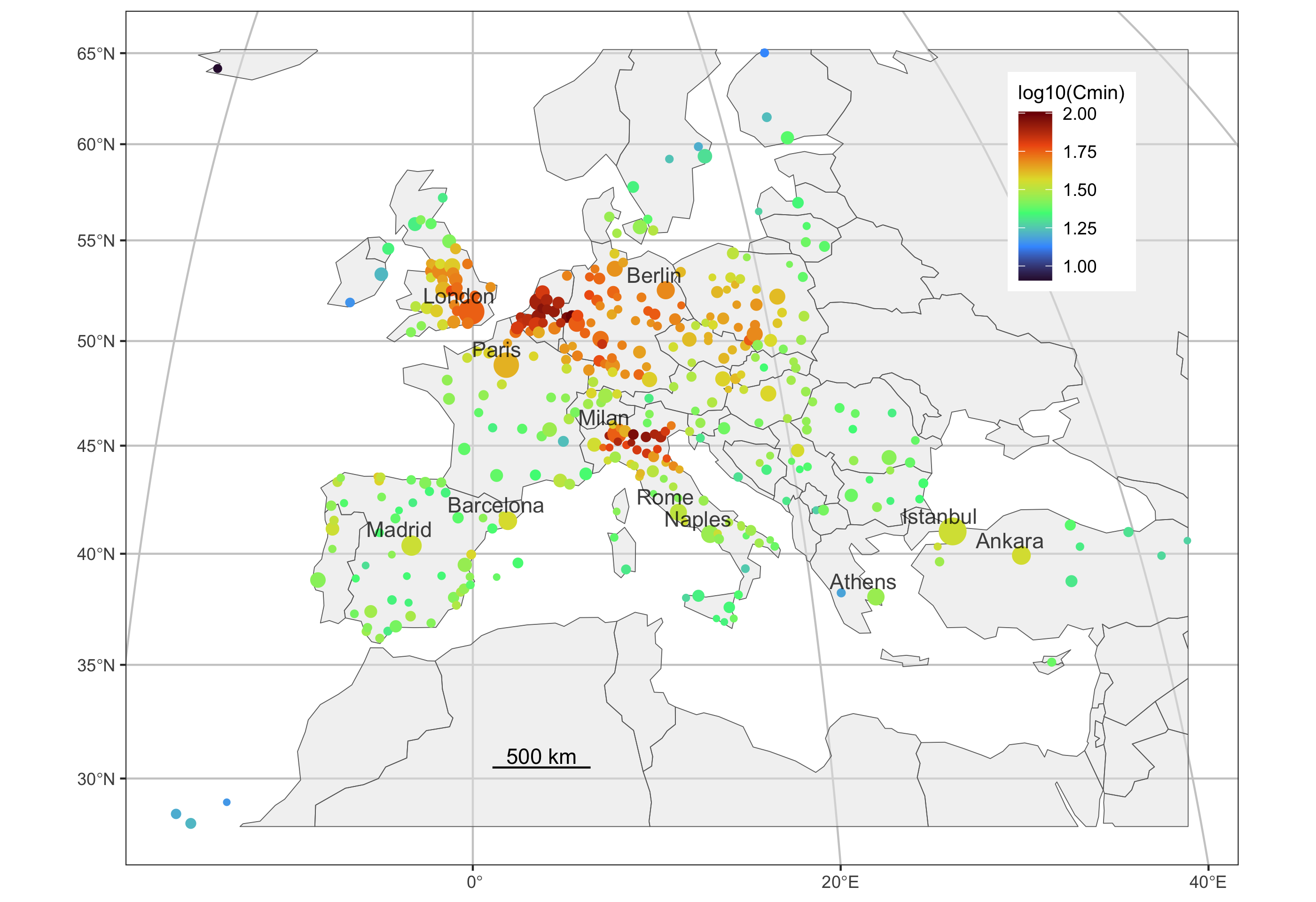} 
\caption{Left: Map of FUAs: symbol proportional to population, colored along size group. Right: Map of NO$_2$ background per FUA. $log_{10}(C_{min})$ from vertical column in $log_{10}(\mu mol /m^2)$}
\label{DataMaps}
\end{figure}

We retrieve total population $N$ from the FUA database. We further define five groups of cities, from Largest (> 5 million) to Smallest (< 300k inhabitants) to later support the display of our results (Figure \ref{DataMaps} left)

\subsection{NO$_2$ data}\label{NO2 data}
We use both annual mean NO$_2$ surface concentrations $G$ and annual mean tropospheric NO$_2$ vertical columns $C$.

Surface concentrations $G$ were collected for the calendar year 2018 (prior to COVID) from the Air Quality e-Reporting (AQ e-Reporting) dataset developed by the \citet{aqeQualityMetaInfo}. $G$ is measured in micrograms per cubic meter ($\mu g/m^3$). The dataset contains records from the monitoring stations of each member country of the European Economic Area. The data for each station includes the annual mean NO$_2$ concentration, verification and validity of the measured values, time and data coverage, local context (mixed, traffic, industrial) in addition to station coordinates from which we compute Euclidean distances $r$ to our defined centers.

We selected verified and valid records with a time and data coverage both greater than 75\% \citep{aqeQualityMetaInfo}. Our subset includes 1,397 monitoring stations of which 674 are in a mixed, 566 in a traffic, and 157 in an industrial context. Even in cases where the number of stations per city is low, the information can be used in our case because cities are pooled together and we do not aim to produce a local prediction map for each city.

Tropospheric NO$_2$ concentrations $C$ were taken from the TROPOspheric Monitoring Instrument (TROPOMI) of the Sentinel-5 Precursor Earth observation satellite. $C$ is measured in micro moles per square meter ($\mu mol /m^2$). We used the Sentinel-5P Pre-Operations Data Hub \citep{sentinel5phub} to collect daily tropospheric NO$_2$ vertical columns for all urban areas and over a whole year from October 25, 2018 to October 24, 2019 . The time period was chosen to be as much as possible in line with the monitoring stations data, knowing some data was not available prior to Oct. 25, 2018 \citep{sp5No2SatelliteReadMe}. Mosaicking and geometric operations were conducted using \citet{sp5No2SatelliteUserManual2019,sp5No2SatelliteReadMe,snapesa} to obtain best same spatial resolution across all cities, i.e. 7km. We computed the Euclidean distance $r$ from the center of each pixel to the defined city centers.

In contrast to an approach where the main objective is to produce a local prediction of NO$_2$ concentration within each city, we are not particularly impacted by the available spatial resolution because we seek a general model and pool all cities together. Others use similar resolution data but downscale pollution to higher spatial resolution using population grid, wind direction and urban characteristics \citep[e.g.][]{iungman2024impact}. In our case it is preferable not to downscale data \textit{ex ante} in order not to force the NO$_2$ internal distributions to follow the density distributions and avoid a circular reasoning. Since we estimate a continuous profile from observed points anyway, we actually perform a local prediction - or downscaling - without imposing such a constraint.

We compute the minimum value of $C$ per FUA, $C_{min}$, in order to capture a background regional value for each FUA. As shown on the map of $C_{min}$ (Figure \ref{DataMaps} right), there is an important regional effect at the level of the continent with high pollution levels along the so-called `blue banana', i.e. the traditional core industrialized areas of Europe from Liverpool to Milan via Benelux and Rhineland. We do not aim to explain this general spatial trend further but to abstract city effects (both size and distance) from this regional background, hence the computation and use of $C_{min}$ as a control.

As demonstrated in the Supplementary Material \ref{PopCmin}, the two aggregate attributes of our urban areas , i.e. population $N$ and the background pollution $C_{min}$,  are independent. Areas of high pollution concentration comprise cities from all size groups.

\subsection{Regressing NO$_2$ concentration on city size and centrality} \label{regressions}
We estimate different specifications of log-log models, i.e. power-laws in accordance with most of the urban scaling literature. Namely, we regress both concentrations log$_{10}$G and log$_{10}$C against the logarithm of both the Euclidean distance from the city center, log$_{10}$r, and population size, log$_{10}$N. In the case of log$_{10}$G, we also use dummies to control for different local contexts of monitoring stations, with mixed contexts used as reference ($T=1$ for traffic and $I=1$ for industry, both being null otherwise). In the case of log$_{10}$C, we include the minimum concentration over each FUA in logarithmic form, log$_{10}$C$_{min}$, to control for regional background levels.

The following complete equations \ref{Gmodel} and \ref{Cmodel}  were estimated using OLS with robust standard errors:
\begin{eqnarray}
\label{Gmodel}
log_{10} G &=& k_0 + a\ log_{10} N + b\ log_{10} r + k_T\ T + k_I\ I \\
\label{Cmodel}
log_{10} C  &=& k_0 + a\ log_{10} N+ b\ log_{10} r + k_{C_{min}} \ log_{10} C_{min}
\end{eqnarray}

\noindent where the parameters to be estimated are $a$ and $b$, our coefficients of interest for city size and centrality respectively, $k_0$ the intercept, $k_T$ and $k_I$ the stations context effects, and $k_{C_{min}}$ the coefficient associated with the regional level concentration.

All data was pre-processed using Python 3.7.2 while regressions and graphics were made with R 4.4.1.

The independent variables were added sequentially and switched on or off to identify potential cross-correlations and demonstrate the importance of a joint estimation of within (distance) and inter-urban (size) effects. Each model was estimated linearly and two-tailed t-tests performed using the robust covariance matrix estimators (\textit{sandwich} R package with clustered covariance (\textit{vcovCL}) set to \textit{HC1}). We also report the relative importance (share of $R^2$) of each variable, that is the partition after averaging variances over all possible orderings of the independent variables (using the \textit{relaimpo} R package with $lmg$ option).

\section{Empirical results}\label{Article2_Results}

Regression results are reported in Table \ref{allcityStation} for ground NO$_2$ $G$ and Table \ref{allcityRS} for tropospheric NO$_2$ $C$. 

In the case of ground concentrations $G$, models 1 and 3 are separate models for the effect of population and distance. Models 2 and 4 add local context effects to both models. Model 5 combines population and distance effects. Model 6 is the complete model including local contextual effects. Similarly in the case of the tropospheric column concentrations $C$, models 7 and 8 are separate models, while model 9 is the combined model and model 10 the full model including the regional background effect. 

\begin{table}[htbp]
\centering
\begin{tabular}{lrllllll}
\toprule
\multicolumn{8}{c}{$log_{10}G$} \\
& & (1)   & (2)   & (3)   & (4)   & (5)   & (6) \\
\midrule
\multicolumn{8}{l}{\textbf{Coefficients}} \\
Intercept & $k_0$ & 0.56*** & 0.48*** & 1.65*** & 1.36***   & 0.73*** & 0.59*** \\
& & (0.06) & (0.05)& (0.04) & (0.03)  & (0.05) & (0.04) \\
& & [9.86] & [10.50]& [44.69]& [43.54] & [13.98] & [13.53] \\

City size $log_{10}N$ & $a$ & 0.14*** & 0.13*** &    &     & 0.22*** & 0.19*** \\
& & (0.01) & (0.01) & & &  (0.01) & (0.01) \\
& & [14.55] & [17.89]& & & [22.26] & [23.71] \\

Distance $log_{10}r$ & $b$ & & & -0.07*** & -0.02*     & -0.18*** & -0.12*** \\
& & & & (0.01) & (0.01)  & (0.01) & (0.01) \\
& &  & & [-7.44]& [-2.34] & [-15.52] & [-13.03] \\
\multicolumn{8}{l}{Context (ref. = mixed)}\\
\ \ traffic $T$ & $k_T$ &    & 0.25***&  & 0.24***  & & 0.22*** \\
& &   & (0.01) &    & (0.01) &    & (0.01) \\
& &   & [29.06] &    & [25.27] &    & [26.87] \\
\ \ industrial $I$ & $k_I$ & & -0.03$^+$ &   & -0.05**  &    & 0.002 \\
& &    & (0.01) &    & (0.02) &    & (0.01) \\
& &   & [-1.83] &    & [-2.81] &    & [0.15] \\
\midrule
\multicolumn{8}{l}{\textbf{Relative importance}} \\
City size $log_{10}N$ &  &    &  0.26    &    && 0.65 & 0.31 \\
Distance $log_{10}r$  &    & & &    & 0.06   & 0.35 & 0.12 \\
Context & &    & 0.74&    & 0.94 &    & 0.58 \\
\midrule
R$^2$ & & 0.13 & 0.47& 0.04 & 0.36  & 0.29 & 0.54 \\
dF& & 1395  & 1393  & 1395  & 1393  & 1394  & 1392\\
\bottomrule
\multicolumn{8}{l}{(std errors) [t-values]}\\
\multicolumn{8}{l}{***$\rho$ $\textless$ 0.001 **$\rho$ $\textless$ 0.01 *$\rho$ $\textless$ 0.05 $^+$$\rho$ $\textless$ 0.1}

\end{tabular}
\caption{log$_{10}$G ground stations NO$_2$ regression models}
\label{allcityStation}
\end{table}%

\begin{table}[htbp]
\centering
\begin{tabular}{lrllll}
\toprule
\multicolumn{6}{c}{$log_{10}C$} \\
& & (7)   & (8)   & (9)   & (10) \\
\midrule
\multicolumn{6}{l}{\textbf{Coefficients}} \\
Intercept & $k_0$ & 0.72*** & 1.71*** & 1.20***  & 0.16***  \\
& & (0.01) & (0.02) & (0.02) & (0.01) \\
& & [58.60] & [96.54] & [74.17] & [18.07] \\

City size $log_{10}N$ & $a$ & 0.16*** & & 0.22*** & 0.14***\\
& & (0.002) &  & (0.002) & (0.001) \\
& & [79.00] & & [96.05] & [99.36]\\

Distance $log_{10}r$ & $b$ &  &-0.01* & -0.18*** & -0.15***\\
& &  &(0.004) & (0.004) & (0.002) \\
& & & [-2.04] & [-44.47] & [-64.22]\\

Background  $log_{10} C_{min}$ & $k_{C_{min}}$ & & & & 0.87***\\
& & & & & (0.003)\\
& & & & & [335.71]\\
\midrule
\multicolumn{6}{l}{\textbf{Relative importance}} \\
City size $log_{10}N$ &   &    & & 0.87 & 0.16 \\
Distance $log_{10}r$  &   &    & & 0.13 & 0.03 \\
Background & &  &  &    & 0.81 \\
\midrule
R$^2$ & & 0.18 & 0.00 & 0.24 & 0.85 \\
dF & & 24814 & 24814 & 24813 & 24234 \\
\bottomrule
\multicolumn{6}{l}{(std errors) [t-values]}\\
\multicolumn{6}{l}{***$\rho$ $\textless$ 0.001 **$\rho$ $\textless$ 0.01 *$\rho$ $\textless$ 0.05 $^+$$\rho$ $\textless$ 0.1}

\end{tabular}

\caption{log$_{10}$C tropospheric column  NO$_2$ regression models}
\label{allcityRS}
\end{table}

Overall, our complete models (column 6 Table \ref{allcityStation} and column 10 Table \ref{allcityRS}), despite their simplicity are very effective to explain NO$_2$ levels, with an $R^2$ of 0.54 for $G$ and 0.84 for $C$ respectively. Differences in the $a$ and $b$ coefficients between the complete models (5 and 6, or 9 and 10) and the partial ones (1 through 4, or 7 and 8) shows the importance of the joint estimate of distance and size effects. The consistency of the coefficients when local or regional controls are added (6 or 10 compared to 5 and 9) and across the two measurement types demonstrates the robustness of our approach.

\subsection{City size and centrality effects}
We find expected signs for both city size and centrality on NO$_2$ concentration: the pollution is significantly higher in larger cities and significantly lower when one moves away from the city center. More precisely, we find that NO$_2$ increases as a power-law with city size $N$, with an exponent (or elasticity in economic terms), ranging from 0.14 to 0.22 depending on the specification of the model and measurement type. It also decreases as a negative power of the distance $r$ from the center, with an exponent ranging from -0.18 to -0.12 (Table \ref{allcityStation}, models 5 and 6, and Table \ref{allcityRS} models 9 and 10).

Population size $N$ is clearly more important than the distance effect $r$, explaining twice as much of the variance and even more when considering tropospheric (satellite) rather than ground data (relative importance in Table \ref{allcityStation} and \ref{allcityRS}).

Importantly, our results are consistent whether we use NO$_2$ concentrations $G$ from ground-based monitoring stations or concentrations $C$ in vertical columns from satellite measurements. Despite differences in metrics, in sampling, and in the controls operated for each of the two data types (local context or regional background), coefficient estimates are in the same range. This is good news especially given the fact that the unconditional correlation between averaged ground and tropospheric concentrations is low (see Supplementary Material \ref{sectcorrGC}).  In addition to bringing robustness to our results, this consistency demonstrates the effectiveness of our approach that looks at both the within and inter city variations simultaneously, rather than aggregating pollution over each city with no account of the centrality, or compared to case studies (see Table \ref{allcityStation} models 1 to 4 and Table \ref{allcityRS} models 7 and 8 for models with only one of those effects). Moreover, by including centrality without a maximum distance cut-off, but a functional definition of urban areas, we indirectly control for potential differences in the spatial extent of cities of similar population size.

\subsection{Local and regional effects}
We also identify important effects from both the local context within a city and the wider regional background.

In the case of concentrations $G$ measured at the ground, the location of monitoring stations in traffic or industrial contexts was compared to reference mixed contexts. Higher values of air pollution (+66\%, i.e. $10^{0.22}$) are found in a traffic context, as expected from the local street effects literature \citep[e.g.][]{Kirby98, doi:10.1515/mgr-2015-0015}.The effect of being situated in an industrial context is not significant, thus emphasizing traffic as the main source of NO$_2$.

In the case of the tropospheric column measurements $C$, we find as expected that the regional background concentration, as measured by the minimum concentration in the urban area $C_{min}$, accounts for most of the variance. Yet controlling for this background does not substantially change population size and centrality effects, which again validates our endeavor. The background effect is also modeled as a power-law (Table \ref{allcityRS}, model 10) and its exponent is 0.87, which is not very far from linearity. NO$_2$ concentrations are almost additive (extensive variables) and thus urban emissions simply add to the wider regional context emissions.

\subsection{Predicted profiles}
We obtain good predicted values for each city from both full models 6 and 10. Mapping predicted values within cities using the ground NO$_2$ model (model 6) would however suffer from the uneven distribution of monitoring stations and would require an interpolation across different, and unknown, local contexts within each city.

Conversely, satellite data cover each urban region evenly, with constant spatial resolution, and the control background parameter applies equally to every pixel of a city. Mapping simplifies in that case to one dimension, that is a profile of NO$_2$ along the distance $r$ for each city. We have seen that the regional background control $C_{min}$ is important (compare model 9 and 10), and that it depends on the relative location of cities within Europe. Yet, city size does not depend on this location within Europe: there are both small and large cities in the most/least polluted wider regions (see Supplementary Material \ref{PopCmin}). Therefore, the results of the satellite based models can effectively be used to show the general effect of distance and size after removing the background effect and after grouping cities by size.

\begin{figure}[htbp]
\centering
\includegraphics[width=0.8\textwidth]{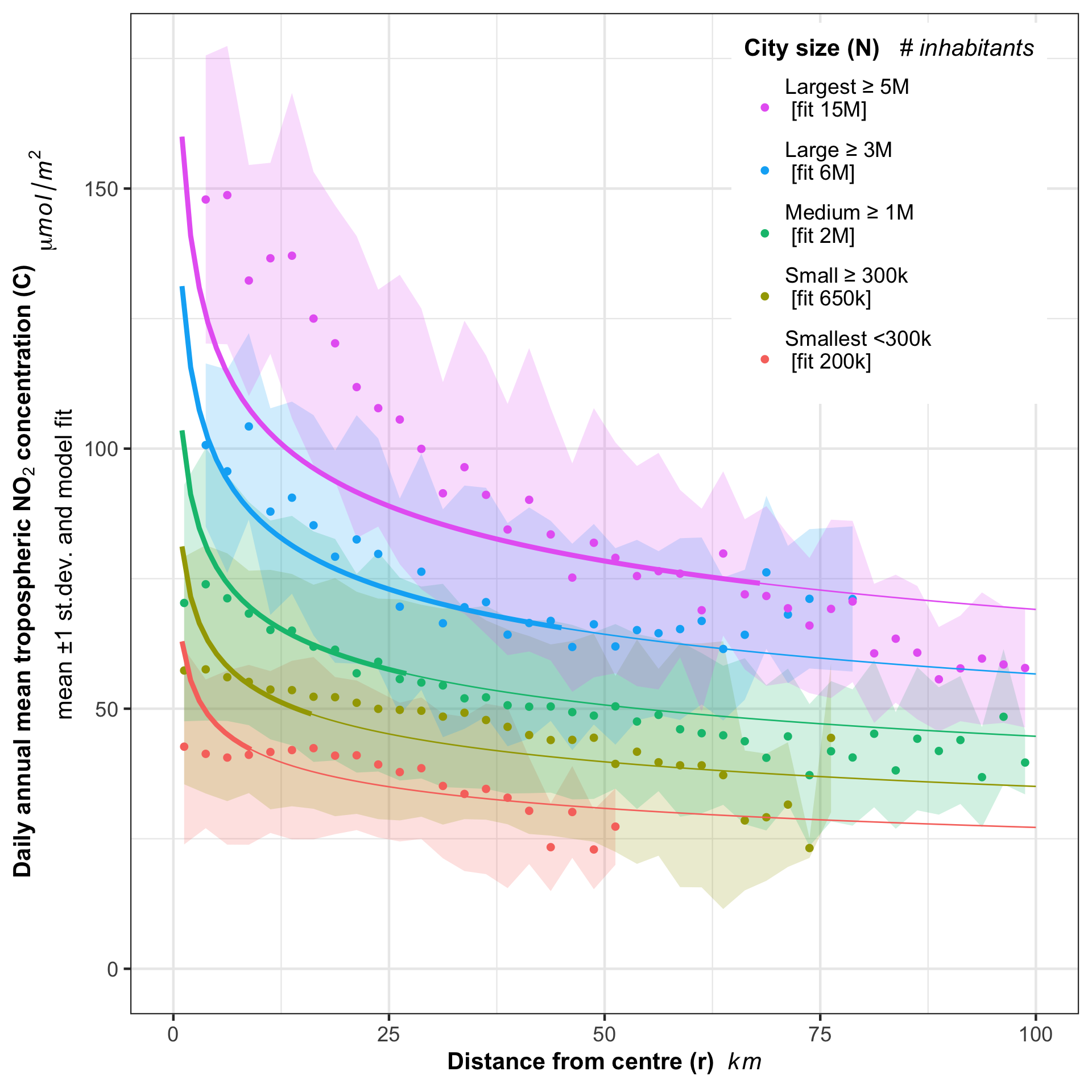} 
\caption{NO$_2$ concentration profiles $C_N(r)$ in European cities (n=378). Dots: mean per distance bands (2.5km) from the center and city size groups. Envelope: $\pm$1 st. dev. Curve: Fitted power-law for representative population of the group ($C_N(r)=10^{1.20} {N^{0.22}}/{r^{0.18}}$)(bold: distance range for a linear scaling of  NO$_2$/capita using $N_{ref}=10^6$ and $d_{ref}=20km$).}
\label{PlotNO2DistGroup}
\end{figure}

Figure \ref{PlotNO2DistGroup}  shows the average NO$_2$ concentrations $C$ ($\pm 1$ standard deviation) by group of city size along with the corresponding model prediction without the background effect, that is with predictions from model 9. The graph thus illustrates both the inter- and intra-urban effects of $N$ and $r$ on NO$_2$ concentrations $C$ irrespective of the background pollution. It shows the decline of concentrations averaged over distance $r$ intervals for our five groups of cities, as well as an estimate for a representative population of each group. Because we display the estimates per group, the observed values at tail are slightly flattened (and lower than the fit) due to the further expansion of the largest cities within each group. Also the fit for the Largest cities group seems a little too low due to London, Milan and Paris being within the high pollution area of Europe. Despite those artificial limits, the graph shows that the estimated power-law is an effective representation of both agglomeration (population size) and centrality effects.

\begin{figure}[htbp]
\centering
\includegraphics[width=0.8\textwidth]{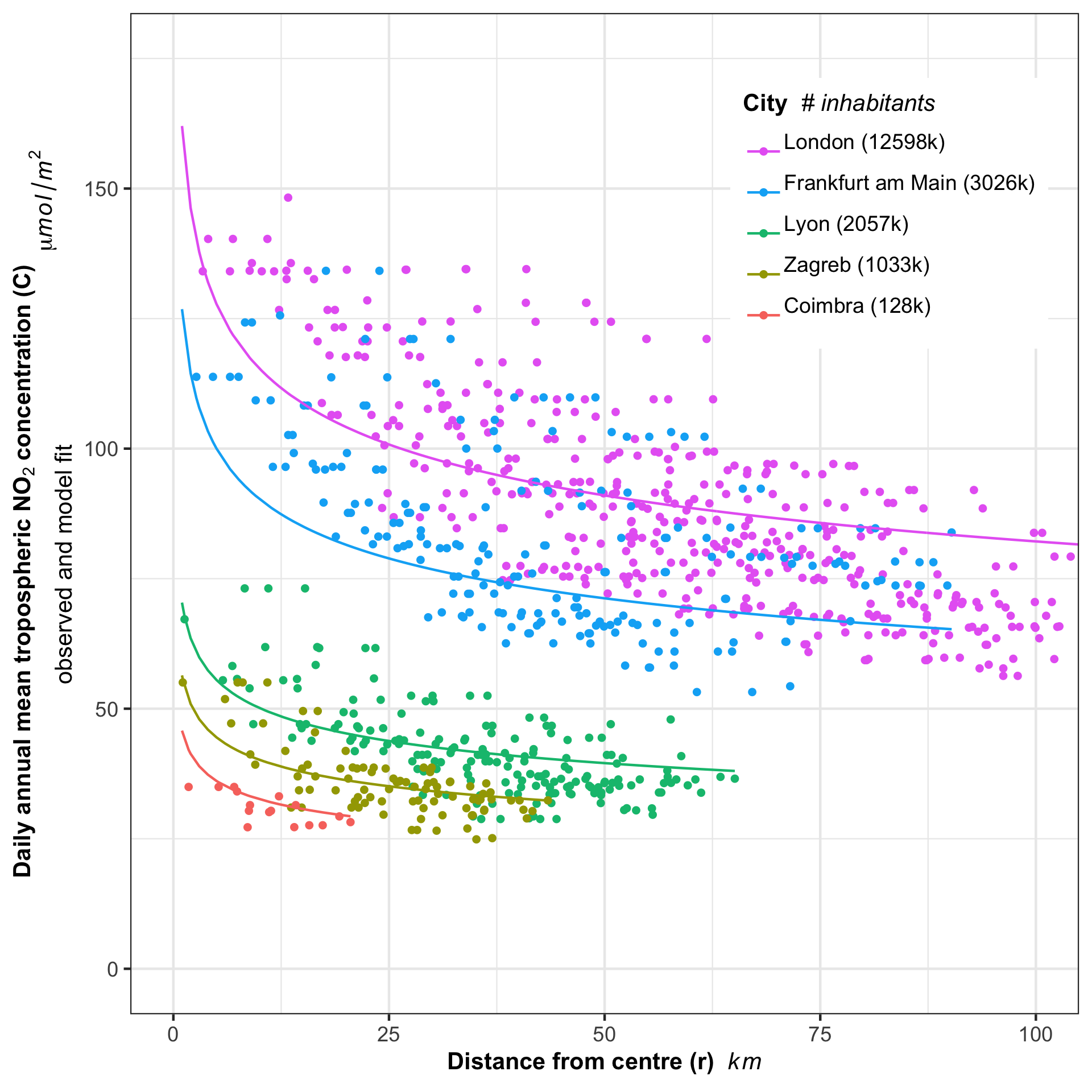} 
\caption{NO$_2$ concentration profiles $C(r)$ for a representative city per group. Dots: observed values per pixel. Curve: Fitted power-law as of model 10, i.e. $C(r)=10^{0.16} {C_{min}}^{0.87} {N^{0.14}}/{r^{0.15}}$.}
\label{Plotspecific}
\end{figure}

Figure \ref{Plotspecific}  shows an example city for each group and the fitted model 10, which therefore re - integrates the city-specific background level $C_{min}$ and the city-specific population $N$. We can see that the individual fits per city are even better (with no tail effect). Despite being very parsimonious, our model predicts quite well how NO$_2$ levels decrease as one moves farther to the periphery of a given city.

\section{Generalisation and discussion} \label{scaling}
In this section, we mathematically generalize the effects of centrality and city size on NO$_2$ concentration in order to understand how the total air pollution over an urban area scales with population size $N$, thereby hinting into per capita pollution at city level. First, we provide a general formulation, then introduce reasonable parameters from our empirical work or the literature, discussing the meaning of these parameters.

\subsection{General formulation}
A general form for the radial profile $C_N(r)$ of NO$_2$ along the distance $r$ from the city center for a city of population $N$ is
\begin{equation}\label{radprof}
C_N(r)=C_1 N^a r^{b},
\end{equation}

As demonstrated by our empirical log-log estimates, this power functional form is valid for both concentration near the ground and concentration in the tropospheric column. For clarity in the rest of the exposé, we will continue to use the $C$ notation, although we could equally use $G$. In this expression, $C_1$ is a constant concentration and, as previously defined, $a$ is the exponent (elasticity) with respect to population size and $b$ the exponent (elasticity) with respect to distance from the center.

In order to compute the total NO$_2$ air pollution, that is the total quantity $P_N$ of NO$_2$ in the air above a city of population $N$, we must integrate the radial distance profile (equation \ref{radprof}) over the whole urban area, that is from the center to the fringe.

The total pollution over a city of size $N$ is
\begin{equation} \label{concentration}
P_N \ =\ \int_0^{r_f}C_N(r)2\pi r \,dr \ = \ 2\pi C_1 N^a\frac{r_f^{2+b}}{2+b},
\end{equation}

where $r_f$ is the radius (extent) of the city.

It is worth noting that this is not associated with an additional hypothesis in the case of concentration in the tropospheric column, since it is measured per unit of surface area in $\mu$mol/m$^2$. Furthermore, the integration resolves a minor theoretical issue pertaining to the use of a negative power law to describe the influence of distance $r$. The function diverges right at the center ($r\rightarrow 0$), which is unrealistic. This issue is resolved by integrating over the entire urban area.

We can take a further step because we know that the radius of the urban fringe, $r_f$, is itself not independent of the population size. It is possible to express the relationship between the radius $r_f$ and the population $N$ of a city in terms of a power law
\begin{equation}\label{rfN}
r_f=r_1 N^c
\end{equation}

where $r_1$ is the (theoretical) radius of a `unitary' city with population $N=1$.

This relation is simply the square root of the relationship between the area $A$ of a city and its population $N$, which is known as the `fundamental urban allometry'  \citep[e.g.][]{RIBEIRO20231}:
\begin{equation}\label{AN}
A\sim N^{2c},
\end{equation}

Including the scaling of $r_f$ into the integrated pollution equation (\ref{concentration}) yields
\begin{equation}\label{total_abc}
P_N=\frac{2 \pi C_1 r_1^{2+b}}{2+b} N^{a+bc+2c}
\end{equation}

whereby the total pollution $P_N$ scales with the city's population size, $N$, to the exponent $a+bc+2c$ .

\subsection{Size and centrality exponents $a$ and $b$}
From our empirical estimates, we know (for both the tropospheric column concentration $C$ and the ground-based concentration $G$) that $a$ is positive and in the range $[0.14,0.22]$, and $b$ is negative and in the range $[-0.18,-0.12]$.

Given these values and their variations across the different specifications, it is reasonable to set $a=-b=1/6 (\simeq 0.16)$ for simplicity. This value of $0.16$ was also obtained for city-size effects in a meta-analysis by \citet{wyf2023}. We can therefore replace equation \ref{radprof} and model the NO$_2$ concentrations profile over any city of size $N$ using

\begin{equation}\label{Nr16}
C(N,r) \sim  (N/r)^{0.16},
\end{equation}

On a log-log scale, this relationship produces a straight line for any given background level. In the Supplementary Material \ref{sectCNr} we show that our simplified model effectively represents the data as soon as the wider regional background pollution is considered separately.

The equation means that on average, a 1\% increase in city size corresponds to a $0.16$\% increase in air pollution, while a 1\% increase in distance from the center corresponds to a $0.16$\% decrease in pollution. Since $10^{0.16}\simeq 1.45$, this also implies that air pollution is 45\% higher in a city with a population 10 times larger. It is also 31\% lower if you move ten times further away from the city center ($10^{-0.16}\simeq 0.69$). If, instead, the population, respectively the distance, is multiplied by a factor 2, the effect on air pollution is +12\% ($2^{0.16} \simeq 1.12$), respectively -10\% ($2^{-0.16} \simeq 0.90$).

An exponent of 0.16 is rather low, meaning the increases and decreases are rather slow. This is quite fortunate for inhabitants of central areas and large cities, who would suffer much more near the city center if the exponent was higher. However, it is also concerning for inhabitants of more peripheral areas, since a low exponent means that the decrease with the distance from the center is also slow. Air pollution diffuses and spreads over large regional areas, as evidenced by the fact that the minimal concentration $C_{min}$ plays such a significant role in our analysis of satellite measurements. This means that people living in urban areas in a very wide sense (not just central urban cores) breathe heavily polluted air.

\subsection{Urban extent and the $c$ exponent}
With $a$ and $b$ now fixed, the result of the integral equation (\ref{total_abc}) also simplifies:
\begin{equation}\label{PN}
P_N\sim N^{1/6+11c/6}
\end{equation}

This scaling law expresses the relationship between total air pollution and population size, depending on $c$, which determines how a city's surface relates to its population. Therefore, the super-linearity or sub-linearity of the total pollution ultimately depends on $c$, which, although unobserved, reflects the way in which we decide to delineate cities, i.e. define their fringes.

In the urban allometry literature where equation (\ref{AN}) is defined and estimated \citep[e.g.][]{BattyFerguson, Rybski_etal2017, RIBEIRO20231}, the exponent $2c$ is known to lie between $2/3$ and $1$. Therefore, our $c$ can potentially take values between $1/3$ and $1/2$. With $c=1/2$ we would assume a linear scaling of the urban surface area as the population grows. Lower values would assume a sub-linear scaling, i.e. that cities become more compact as they grow.

With equation \ref{PN}, we see that a linear scaling of total NO$_2$ pollution with population would be obtained with $1/6+11c/6=1$, that is with $c = 5/11 \simeq 0.45 $. For higher (respectively lower) values of $c$, total NO$_2$ would be super (sub-)linear, i.e. the total pollution grows more (less) quickly than the population.

A number of authors have recently emphasized the importance of the city definition in the scaling literature and suggested different delineation methods  \citep{ doi:10.1177/2399808318755146, Dong2024}.  \citet{doi:10.1068/b4105c}  showed that a sub-linear or a super-linear relation can be obtained for CO$_2$ emissions depending on the definition. \citet{mennicken2024road} showed that the scaling of road detours (which can also relate to pollution) depends directly on whether city cores or extended urban regions are considered.

We can clearly see again here, thanks to the fact that we have estimated distance profiles, that the scaling behavior is largely a matter of where one decides to trace the external limit of a city.

Using the scaling law exponents of land use ($1/2$) or density profiles ($1/3$), as identified by \citet{LemoyCaruso2020}, it follows that, when the city definition is linked to land use - e.g. its external limit is defined by a minimum proportion of urbanized land -  $c$ will be close to $1/2 > 5/11$, and a super-linear pollution will be found ($1/6+11c/6=13/12\simeq1.08$). If the population density is rather the variable of interest and a minimum threshold is used to define the external fringe of cities, then $c$ will be close to $1/3 < 5/11$ and the scaling of the pollution will be sub-linear ($1/6+11c/6=7/9\simeq0.78$).

\subsection{Pollution per capita }

In order to compute numerical effects for cities of different sizes, it is necessary to use the full equation \ref{total_abc} and identify the other parameters ($a$, $b$ being set and $c$ discussed).

\begin{figure}[htbp]
\centering
\includegraphics[width=0.8\textwidth]{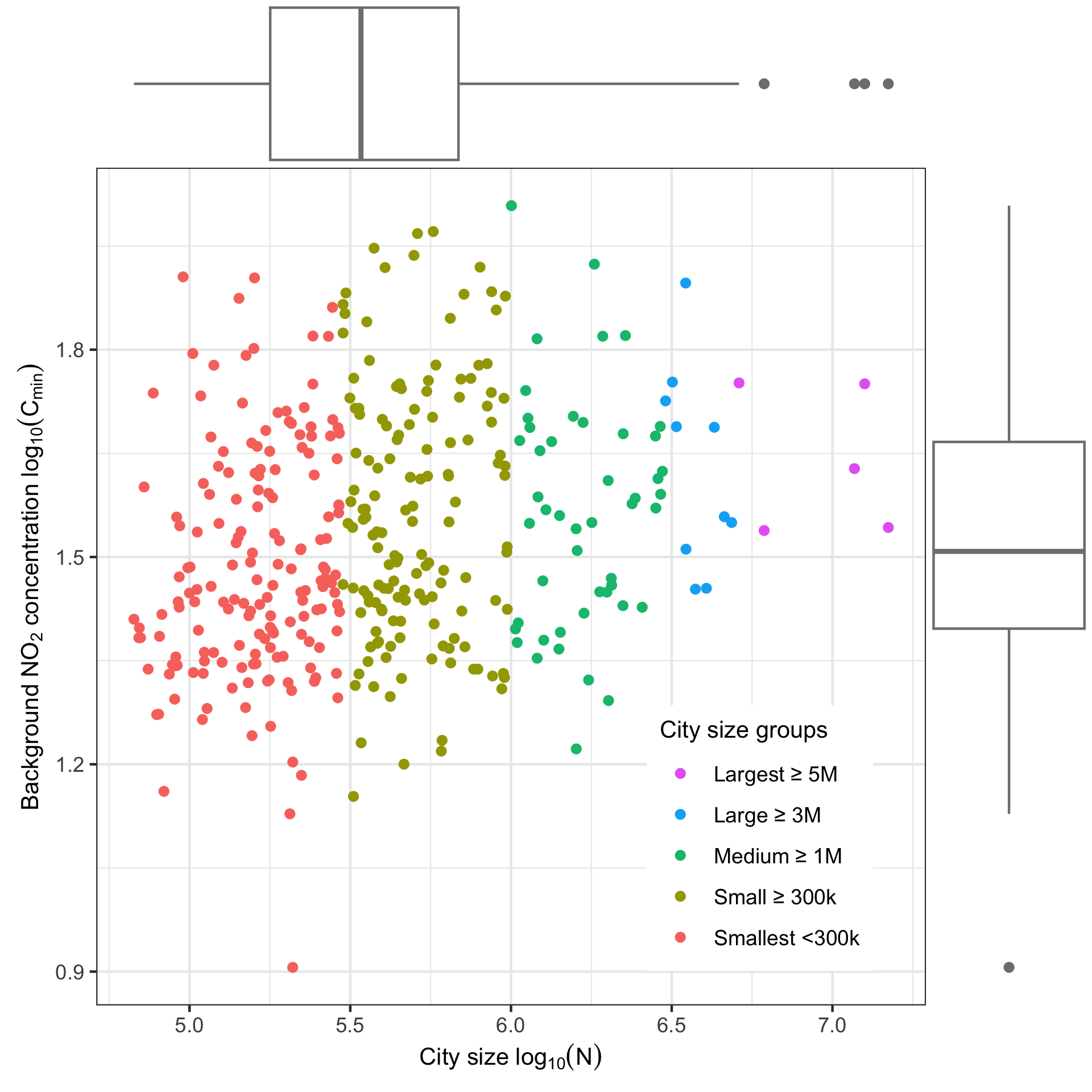} 
\caption{Plot of NO$_2$ background $log_{10}(C_{min})$ from vertical column in $log_{10}(\mu mol /m^2)$ against population size $log_{10}N$. Respective boxplots.}
\label{PlotPopCmin}
\end{figure}

The constant concentration $C_1$ is given by the intercept of our model 9, i.e. the model where both distance and size were included, thus a value of $C_1= 10^{k_{0}}=10^{1.20}\simeq 15.8 \mu mol/m^2$. However, because our $1/6$ simplification is slightly lower than the absolute terms estimates in model 9 for both distance and population effects (but slightly higher than model 10 estimates with background controls), the NO$_2$ concentration at the center would be slightly underestimated. We therefore move a little from the centre for the calibration and use $C_1= 10^{k_{0}}=10^{1.47}\simeq 29.5 \mu mol/m^2$, which is obtained by setting to zero the underestimation at 1km for a city of 1 million inhabitants. $k_{0}=1.47$ is also close to the unconditional median of $log_{10}(C_{min})$ across all cities (see vertical boxplot on Figure \ref{PlotPopCmin}). Using the molar mass of NO$_2$ ($46g/mol$), we can finally express $C1$ in mass concentration to facilitate interpretation: $C_1\simeq 1358 \mu g/m^2$.
 
For the city extent - since a city of a single person do not exist and $r_1$ is unobservable - we consider $r_f$ (equation \ref{rfN}) for a reference city population of 1 million inhabitants, $N_{ref}=10^6$. Let us first assume that such a city is reasonably delineated by a fringe radius of $d_{ref}=20 km$ from the center. For simplicity, we then choose the city definition (fundamental allometry) that yields a linear scaling of total pollution with city size, that is $c=5/11$. This means that for a city of size $N$, the fringe is located at $r_f=d_{ref}(N/N_{ref})^{5/11}$, which is around $60 km$ for a city of 10 million inhabitants such as Paris or London. Note that the end of the bold part of the curves on figure \ref{PlotNO2DistGroup} visually materializes the extent of each representative city under this assumption ($10^6$ city being defined by $20km$). Because of our linear scaling assumption, integrating pollution until that $r_f$ distance thus holds the same pollution per capita, irrespective of city size.
 
All parameters being set, we get a per capita air pollution of $P_1=2 \pi C_1 (d_{ref}N_{ref}^{-c})^{2+b}\simeq 3.57g$, regardless of city size. This is typically produced by a 30km trip with a diesel car (assuming a $122mg/km$ emission factor from the average scenario of \citet{degraeuwe2016impact}, using 0.35 NO$_X$ $g/km$ and a 0.35 NO$_2$ fraction). Since NO$_2$ is removed from the air within about a day \citep[(in winter][]{beirle2003weekly},  through photodissociation and depending on wind and turbulences \citep{roman2023wind}, this can be equated to a day worth of emissions for one individual. A 30 km trip per person, knowing our cities are functional urban areas defined by commuting, seems very sound hence $3.57g$ is a reasonable average across many cities.

What if we had chosen to define the extent of a city of $N_{ref}=10^6$ inhabitants more largely using $d_{ref}=30 km$ or more restrictively, using $d_{ref}=10 km$? In other words, what if we had chosen $r_f\simeq 90 km$, respectively $r_f\simeq 30 km$,  for a city of $N\simeq 10^7$ inhabitants such as Paris or London? These would lead to per capita air pollution of $P_1\simeq 7.5 g$, respectively $P_1\simeq 1g$. The range of those values shows very well the importance of properly defining urban areas depending on research objectives or depending on the envisaged policy instruments, i.e. whether they target city cores only (low level of per capita pollution emitted), or can include farther peripheries (higher level).

\subsection{Scaling of the profiles}
Finally, we can further contribute to urban scaling theory. Since we obtain a power-law expression for the concentration $C_N(r)=C_1 N^a r^{b}$, we can use the fact that power functions are scale-free to identify an isometric scaling of radial concentration profiles.

Equation \ref{concentration} can be rewritten as $C_N(r)=C_1 N^\gamma (r/N^\gamma)^{b}$, with $\gamma=a/(1-b)=1/7\simeq 0.14$, after assuming $a=-b=1/6$. This means that the radial profiles of NO$_2$ concentration have an isometric scaling in 3 dimensions (level of pollution, plus the two spatial dimensions) with city size $N$ to an exponent of approximately $1/7$. This isometry indicates that the pollution profiles of cities of varying sizes are homotheties of each other, which we can see as nested cones (or volcanoes). A similar behavior was found for other profiles, including population density, with an exponent $\gamma=1/3$ \citep{LemoyCaruso2020} or housing price, with an exponent $\gamma=1/5$ \citep{laziou2024radial}.

\section{Conclusion}\label{Concl}
We have conducted the first quantified and simultaneous examination of the effect of both city size and centrality on NO$_2$ for a large set of cities. By doing so, we have addressed a research gap by linking a critical spatial dimension of intra-urban pollution to its inter-urban variations.

We have estimated the profile of pollution along the distance to the main center of an urban area and have characterized how this profile is translated upwards when cities are more populated. Our estimates show for example that, for a given regional or local background, doubling the distance to the center reduces NO$_2$ exposure by about 10\% but doubling the population of a city leads to a 12\% increase. These results question urbanisation and densification strategies as well as the distribution of population across cities.  

To generalize our findings and further link the two scales of analysis, we have seen that both centrality and city size effects can satisfactorily be described using a single exponent of $1/6$, which is in the range of our estimates across different empirical specifications and across our two types of measurements (satellite or ground based). Additional benefits have arisen from the integration of this simplified expression over all distances within a city, to obtain $P_N\sim N^{1/6+11c/6}$, the total quantity of NO$_2$ for a city of given size $N$.

First, our estimates translate into a $3.57g$ NO$_2$ production per person across the entire urban system, or a 30km car trip on average. We believe that such a translation in `car-km' should raise awareness among stakeholders and citizens about air pollution exposures and the level of behavioral or technological change needed. Within cities especially, because of their density, car trips can more easily be shortened and shifted to electric energy, while shorter car trips can be replaced by (electric) bikes and longer ones by transit systems. Given the levels of NO$_2$ concentrations and their radial profiles, every effort must be taken in these directions in both small and large cities.

Second, we have seen how the result of this integral $P_N$ depends on the extent of the city considered, that is the distance $r_f$ until which the sum is computed, and how the extent of urban areas evolves with population size ($c$ parameter). 

Properly appreciating the scaling with population size of emissions is fundamental to decide on how and where, in an increasingly urban world, population is best allocated. Compared to the classical scaling literature that considers a single average per city, we have demonstrated here that adding an intra-urban analysis is key. More specifically, rather than making a strict exogenous choice about any delineation of cities, our joint assessment of centrality and city size scaling permits an examination in continuous terms and therefore the identification of ranges within which a particular scaling behavior holds. Many scaling exponents for aggregate environmental or social outcomes are close to 1 and a systematic examination of their internal distribution, from centers to peripheries of cities, may well lead to opposite findings, depending on where the fringe is set.

In our NO$_2$ case, we have seen that linearity is still in the likelihood range given possible values of the exponent $c$ of the fundamental allometry, and the fact that more extreme values could be taken into the scaling exponent ($1/6+11c/6$)  based on different model specifications.

Given our new understanding of expected pollution levels at varying distances from the main activity centers, regardless of population size, future empirical research could explore the relationship between pollution and its sources, especially traffic volumes. This would help identifying other systematic sources of heterogeneity in pollution profiles, such as the presence of transit systems. However, as of now, comprehensive traffic volume data (not merely simulated data based on population) and modal split data are lacking for a broad range of cities. Consequently, we cannot yet fully articulate variations around the estimated model in terms of sustainable transport and infrastructure policy efficiency.

\bibliographystyle{plainnat}  
\bibliography{NO2_dist_size}    

\newpage
\nolinenumbers
\beginsupplement
\section*{SUPPLEMENTARY MATERIAL}

This is supplementary information related to the article entitled

\textbf{Population size and centrality effects on NO$_2$ air pollution across and within European cities}

\newpage

\section{Population and NO$_2$ background ($C_{min}$)}\label{PopCmin}

\newpage

\section{Correlation between ground $G$ and tropospheric $C$ urban NO$_2$ concentration}\label{sectcorrGC}

NO$_2$ concentrations from monitoring stations points on the ground ($G$) were overlayed with the NO$_2$ concentrations pixels of the tropospheric measurements ($C$) in order to identify the strength of the relation.  Measures were taken at each overlapping instance and the mean of both $C$ and $G$ computed for each city ($n=378$).

\begin{figure}[htbp]
\centering
\includegraphics[width=0.8\textwidth]{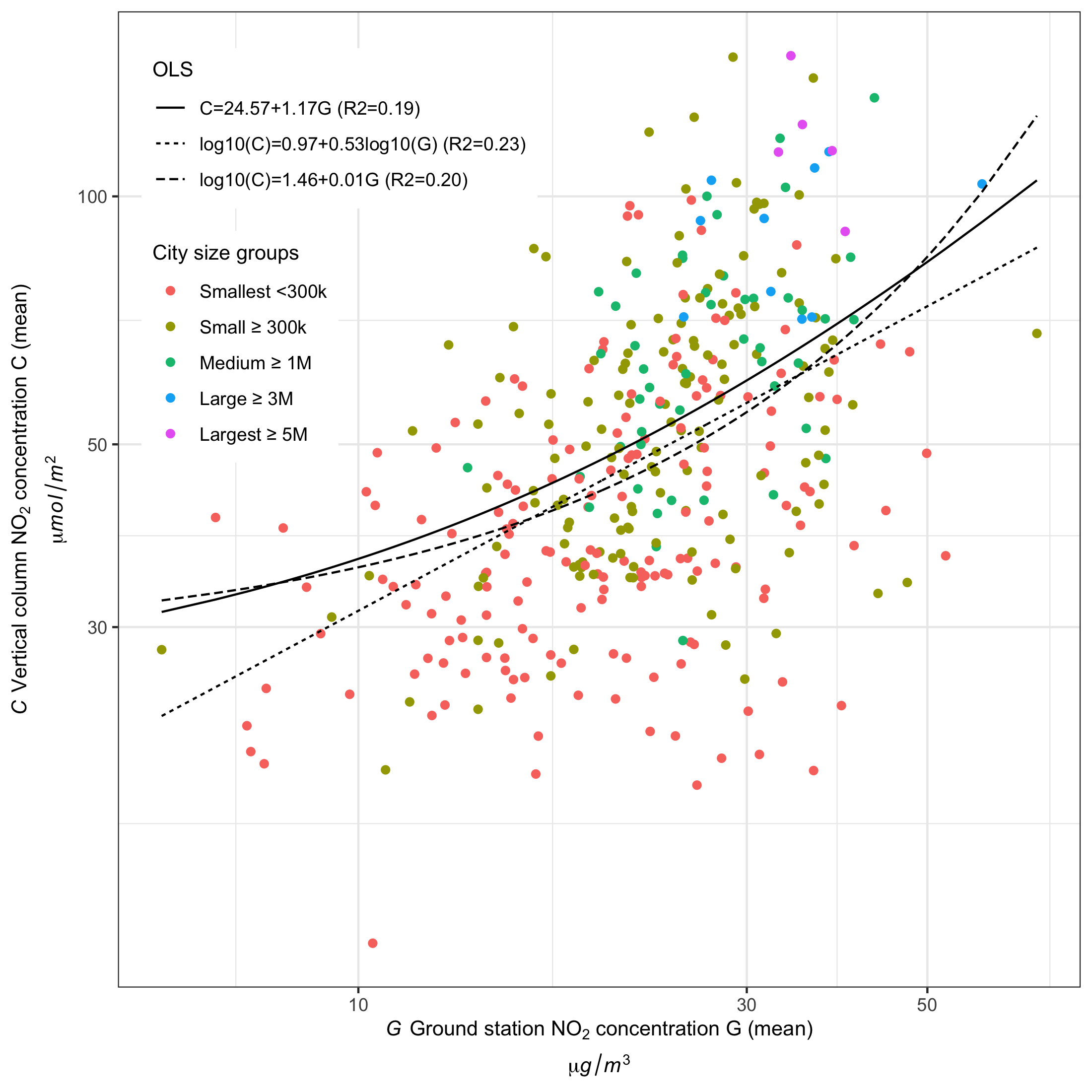} 
\caption{Plot of NO$_2$ tropospheric concentration ($C$) against ground concentration ($G$). Mean over overlapping station and pixel per city.}
\label{PlotGC}
\end{figure}

As demonstrated (with logged axes) on figure \ref{PlotGC}, there is a positive but weak relationship. Linear, log-linear and double log estimates were performed showing a significant ($p<0.001$) positive link, but a low R square, in the range of 0.19 to 0.23. Population size classes are indicated, showing higher levels for larger cities for both measures (top right quadrant) but no specific dispersion along either of the axes.

\newpage

\section{Simplified model validation using $C(N/r)$ (log-log scale)}
\label{sectCNr}

We validate our simplified model $C(N,r) \sim (N/r)^{0.16}$ by plotting $C$ versus $N/r$ on a log-log scale and verifying that a straight line with a slope close to $0.16$ is a good representation of the data. Since the simplified model does not include background levels, we must add it while estimating the equation so that a vertical translation of  $C(N,r)$ is obtained for different background pollution levels ($C_{min}$)

The OLS estimate results in the following equation: $$log(C) = 0.1518\ log(N/r) + 0.009\ C_{min} + 1.0893$$ with all coefficients being significant ($p<0.001$) and the model adjusted $R^2=0.81$.

\begin{figure}[htbp]
\centering
\includegraphics[width=0.8\textwidth]{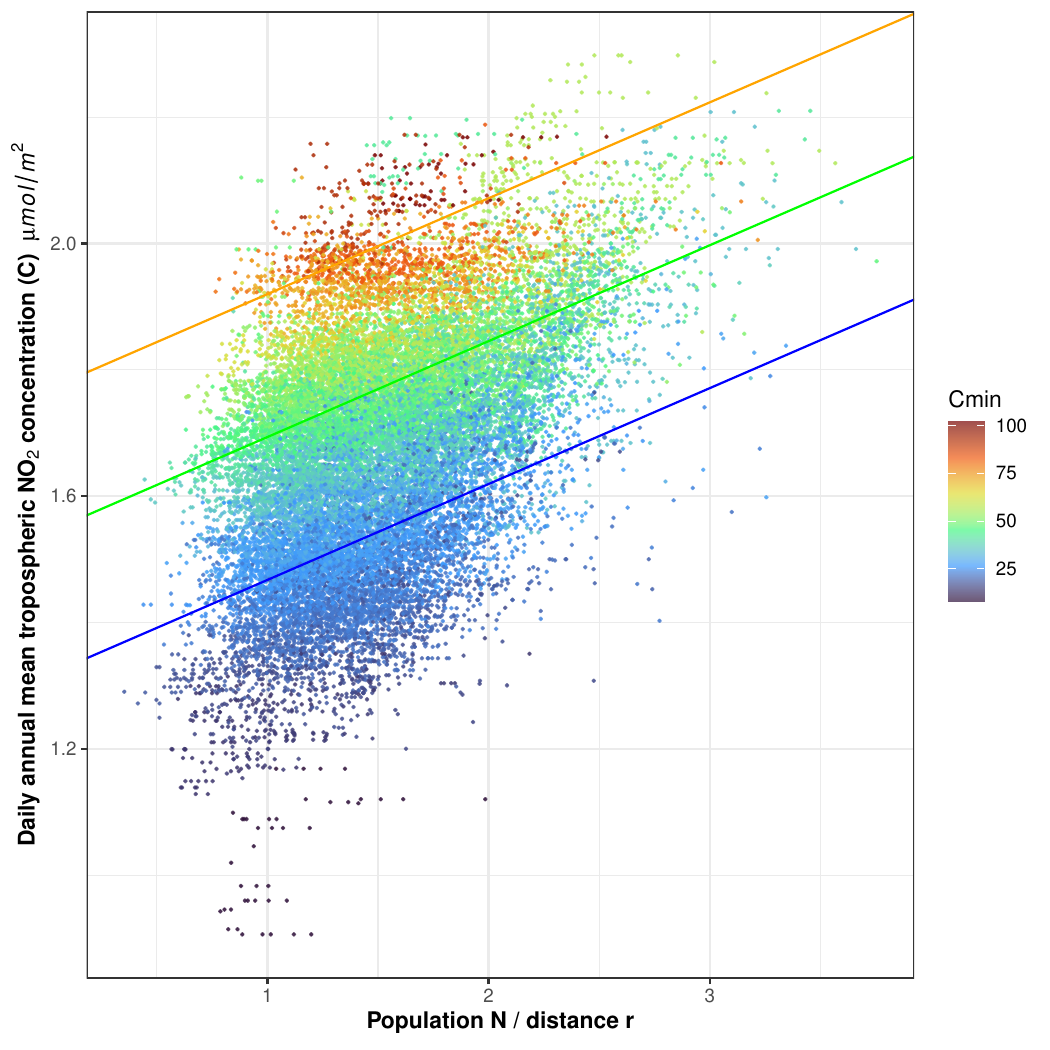} 
\caption{log-log plot of NO$_2$ tropospheric concentration as a function of $N/r$. Colored with background level $C_{min}$ and linear fit for $C_{min}={25,50,75}$}
\label{PlotCNr}
\end{figure}

We thus verify that the slope of this specification ($0.1518$) is very close to our chosen value of $0.16$. In addition, we simulate the straight line for values of $C_{min}={25,50,75}$, respectively in blue, green and orange on Figure \ref{PlotCNr}.

\begin{figure}[htbp]
\centering
\includegraphics[width=0.8\textwidth]{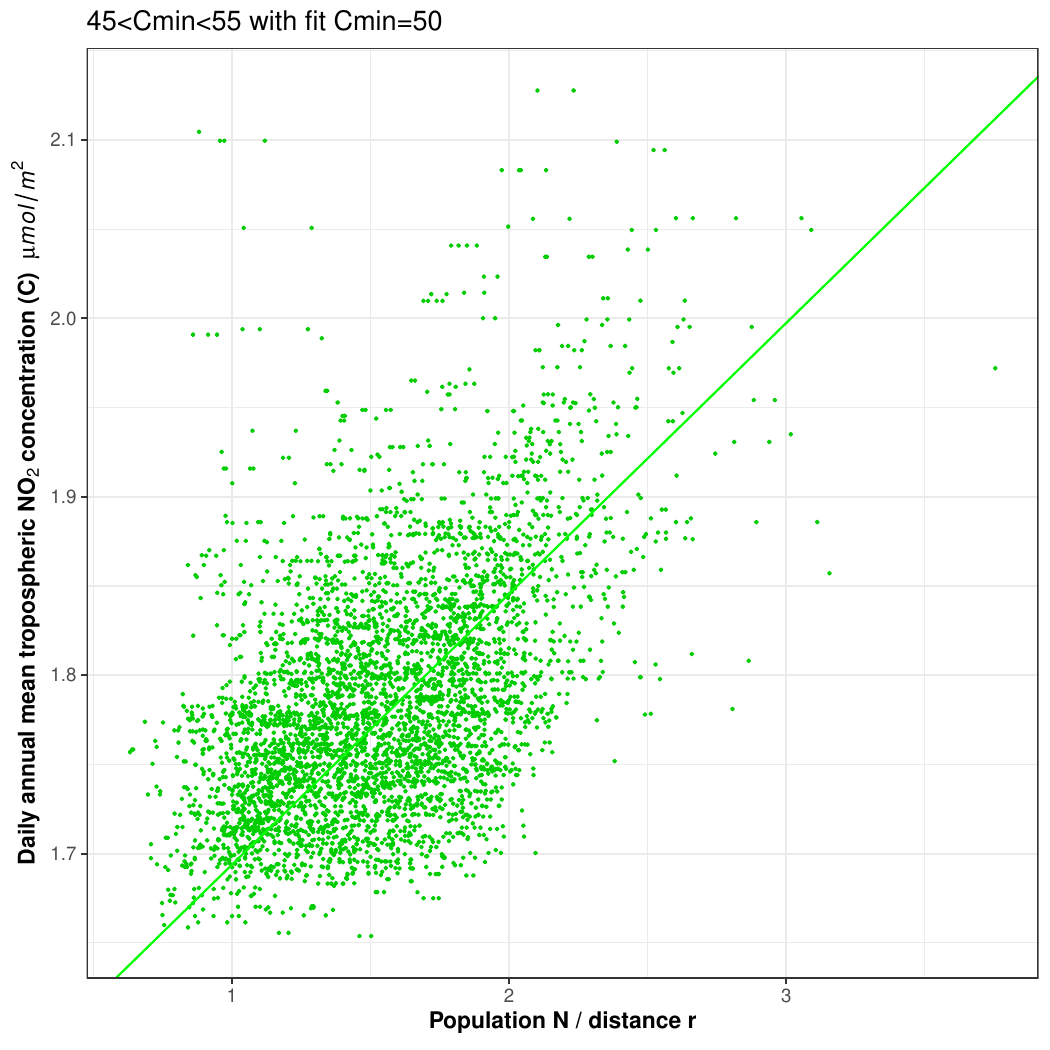} 
\caption{log-log plot of NO$_2$ tropospheric concentration as a function of $N/r$ for background levels $45<C_{min}<55$ and linear fit}
\label{PlotCNr50}
\end{figure}

We see that the point data, taken “color by color” (i.e. for a given range of $C_{min}$) is well represented by the fitted curve with similar color. For clarity, we also provide on Figure \ref{PlotCNr50} the green line separately with the green points, i.e. for a range of $C_{min}$ around $50$ .

\end{document}